\documentclass{iopjournal}

\usepackage{dcolumn}
\usepackage{bm}
\usepackage{braket}
\usepackage{amsmath}
\usepackage{amssymb}
\usepackage{lineno}
\usepackage{mathtools,leftindex,tensor,mhchem}
\usepackage{color,soul}
\usepackage{cite}
\usepackage{float,needspace}
\usepackage{xparse}
\RenewDocumentCommand{\leftindex}{ O{} m }{%
  \mathrlap{_{#1}}#2%
}
\hypersetup{colorlinks=true,urlcolor=blue,citecolor=blue,linkcolor=blue}
\usepackage{graphicx}

\newcommand{\magenta}[1]{\textcolor{magenta}{#1}}

\begin{document}

\articletype{Paper} 

\title{Integrated Generation and Purification of Entangled Coherent States for Non-Gaussian Teleportation}
\author{Ananga Mohan Datta$^{1,*}$\orcid{0009-0009-2708-8712}, William J. Munro$^1$\orcid{0000-0003-1835-2250}, Nicolò Lo Piparo$^{1}$\orcid{0000-0002-0905-9485} and Kae Nemoto$^{1,2}$}

\affil{$^1$Okinawa Institute of Science and Technology Graduate University, Onna-son, Okinawa 904-0495, Japan} \\
\affil{$^2$National Institute of Informatics, 2-1-2 Hitotsubashi, Chiyoda-ku, Tokyo 101-8430, Japan} \\
\affil{$^*$Author to whom any correspondence should be addressed.}\\
\email{ananga-datta@oist.jp}

\keywords{entangled coherent states, two-mode squeezed vacuum states, purification, teleportation}

\begin{abstract}
Entangled coherent states (ECS) provide a powerful non-Gaussian resource for continuous-variable (CV) quantum communication, but their generation in scalable architectures remains challenging. We propose an integrated photonic scheme that creates high-fidelity quasi-entangled coherent states (quasi-ECS) from a two-mode squeezed vacuum state (TMSVS) via photon subtraction in a symmetric waveguide trimer. The resulting non-Gaussian entanglement is further enhanced by single-photon catalysis, which purifies the distributed state after transmission through lossy channels. The improvement in conditional teleportation fidelity is accompanied by a reduced heralding probability. Using these purified quasi-ECS resources, we analyze a photon-number–based teleportation protocol and demonstrate high-fidelity transfer of both coherent states and Schrödinger cat states. For the logical cat qubit, we use an orthogonal even/odd-cat basis, measurement-dependent phase corrections, and the success-probability-weighted fidelity over six logical input states. For completeness and benchmarking, we also study the Braunstein–Kimble (BK) teleportation protocol using Gaussian TMSVSs. In particular, the teleportation fidelity for cat states exceeds the benchmark value of $2/3$ over a broad range of realistic channel and squeezing parameters. By contrast, the underlying TMSVS performs poorly within the photon-number–based protocol, whereas in the BK protocol it exceeds this benchmark only in a restricted regime of high squeezing and high transmissivity. Our results show that integrated photon subtraction and catalysis enable practical, chip-compatible generation of non-Gaussian entanglement suitable for advanced quantum teleportation and CV quantum networks.
\end{abstract}

\section{Introduction}\label{sec:Introduction}

Quantum teleportation is a key primitive for quantum communication and distributed quantum information processing~\cite{Bennett1993, Braunstein1998}, and has been demonstrated in both discrete-variable (DV)~\cite{Bouwmeester1997} and continuous-variable (CV) platforms~\cite{Furusawa1998}. In CV systems, high-fidelity teleportation of Gaussian states—such as coherent~\cite{Zhao2023} and squeezed states~\cite{Takei:PRL:2005}—is routinely achieved using two-mode squeezed vacuum states (TMSVSs) as entangled resources. However, Gaussian resources are fundamentally limited when the input states are non-Gaussian~\cite{Filip:PRA:2010}, such as Schrödinger cat states, which exhibit Wigner-function negativity and higher-order quadrature correlations~\cite{Ourjoumtsev:Nature:2007}. 
Owing to their nonclassical features, they play a central role in bosonic quantum computation~\cite{Ralph:PRA:2003, Lund:PRL:2008}, quantum information processing~\cite{A_Gilchrist_2004}, and quantum-enhanced metrology~\magenta{\cite{Zhang:Metrology:PRA:2025}}. Experimentally, cat states can be generated from single-mode squeezed vacuum states using linear optical elements~\cite{Asavanant:17, Takase:PRA:2021}, as well as in a variety of physical platforms, including atomic systems~\cite{Omran:Science:2019} and through nonlinear optical interactions~\cite{gorshenin2025preparation}. Significant progress has also been achieved in increasing the amplitude of such states~\cite{Sychev:NP:2017, Wang:25}, enhancing their near-orthogonality for coherent-state qubit encodings.
Despite these advances, the teleportation of such non-Gaussian states using purely Gaussian entangled resources remains fundamentally limited. In particular, at finite squeezing and with transmission loss, Gaussian teleportation protocols can degrade the Wigner-function negativity and other nonclassical features of cat states, constraining the achievable fidelity for inputs with strong non-Gaussian structure~\cite{Lee:Science:2011}.
This limitation has motivated extensive efforts to develop non-Gaussian entangled resources capable of transporting nonclassical structure.

Non-Gaussianity can be introduced by conditional operations such as photon subtraction, photon addition, or photon catalysis~\cite{Dakna:PRA:1997, Birrittella:18, Magana-Loaiza:NQI:2019, Fadrny:npj:2024}. These operations enhance entanglement, produce negativity, and have been applied to improve teleportation performance beyond the Gaussian regime~\cite{Takahashi:NP:2010, Opatrny:PRA:2000}. Among the non-Gaussian resources considered to date, entangled coherent states (ECS) are particularly attractive~\cite{Sanders_2012, Murno:PRA:2000}. 
ECS behave as CV analogues of Bell
states, provide strong nonclassical correlations between spatially separated modes, and support
logical qubit encodings in the coherent-state basis~\cite{Jeong:PRA:2001}.
For sufficiently large amplitudes, their quasi-orthogonal nature enables near-deterministic Bell-state measurements using only linear optics, making them valuable for quantum teleportation~\cite{vanEnk2001,Wang:PRA:2001}, quantum repeaters~\cite{Sangouard:10}, metrology~\cite{Joo:PRL:2011}, and bosonic quantum computation~\cite{Ralph:PRA:2003}. Despite these advantages, generating high-fidelity ECS remains experimentally demanding: existing schemes rely on precise interferometric control, strong nonlinearities, or bulk optical implementations that are difficult to scale and sensitive to loss~\cite{Asavanant:17, Nicola:PRL:2020}.

Integrated photonics offers an attractive path toward scalable non-Gaussian state engineering~\cite{Datta:24, Datta:PRA:2025}. Waveguide arrays and multiport interferometers can implement stable, chip-based linear-optical transformations with low phase drift, while photon-number operations can be realized through heralded detection~\cite{Politi:Science:2008, Meany:LPR:2015}. Advances in large-scale integrated photonic architectures highlight the feasibility of on-chip generation and manipulation of complex nonclassical states, including Gottesman–Kitaev–Preskill (GKP) qubits~\cite{Larsen:Nature:2025}, Greenberger–Horne–Zeilinger (GHZ) states~\cite{Lee:APL:2024, Pont2024}, and arbitrary spatial photonic states~\cite{Zelaya:25}.
Recent progress has shown that parity-selective photon subtraction in multimode integrated waveguides can generate non-Gaussian states with enhanced controllability~\cite{Datta:PRA:2025}. 
These developments suggest the possibility of producing quasi-ECS correlations directly in an integrated architecture, but a complete protocol combining state generation, transmission, purification, and application to non-Gaussian teleportation has yet to be demonstrated.

In this work, we propose a fully integrated scheme for generating and utilizing non-Gaussian entanglement suitable for teleporting Schrödinger cat states. Starting from a TMSVS, we use an
integrated waveguide trimer to implement symmetric mixing of the two modes, such that photon
subtraction in the central waveguide erases which-path information and projects the outer modes
onto approximate entangled coherent states.  
Here, we consider symmetric two-mode ECS of the form $\ket{\psi_{\rm ECS}^{\pm}} = \mathcal{N}_{\alpha}^{\pm} ( \ket{\alpha}\ket{\alpha} \pm \ket{-\alpha}\ket{-\alpha})$, where $\mathcal{N}_{\alpha}^{\pm}$ is a normalization constant and the $+$ ($-$) sign corresponds to the even (odd) ECS, respectively. Throughout this work, the photon-subtracted trimer output states are referred to as quasi-ECS to distinguish them from the ECS defined above.
We show that these quasi-ECS attain high fidelity with ideal ECS for realistic squeezing levels.
After propagation through lossy channels, the distributed states are purified using single-photon catalysis implemented via low-transmissivity directional couplers~\cite{Zhang:PRA:2025, lvovsky2002quantum}. This purification step improves the resource fidelity across a broad range of channel parameters, although the corresponding purity improvement depends on the operating point, while remaining compatible with integrated-photonics technology.

Using the resulting purified quasi-ECS, we analyze a photon-number–based teleportation protocol~\cite{vanEnk2001} and evaluate its performance for both coherent states and Schrödinger cat states. We find that coherent-state teleportation attains high fidelity consistent with standard CV results, providing a baseline validation of the resource. More importantly, we show that teleportation of cat states with amplitude $\beta \approx 0.8$ exceeds the standard qubit teleportation fidelity benchmark of $2/3$, applied here to the orthogonal even/odd-cat logical-qubit subspace~\cite{Sandau:PRL:1994, Sandau:PRL:1995}. For CV teleportation of non-Gaussian states, this value is widely used as a reference for assessing teleportation performance~\cite{Caves:PRL:2004}.
In contrast, within the squeezing and channel-transmissivity ranges considered, the TMSVS remains below this benchmark when used with the same restricted photon-number measurement and correction procedure. These results highlight the operational advantage of the trimer-generated quasi-ECS resource within the photon-number-based teleportation architecture, demonstrating its ability to achieve genuinely cat-state teleportation in parameter regimes where the corresponding Gaussian TMSVS resource is ineffective.

 The remainder of this paper is organized as follows. Section \ref{sec:overview} presents an overview of the entanglement-distribution and teleportation scheme. Section \ref{sec:state_generation} analyzes quasi-ECS generation in the waveguide trimer. Section \ref{sec:loss_purification} introduces the channel-loss model and purification protocol. Section \ref{sec:teleportation} evaluates the teleportation performance for cat states. Section \ref{sec:conclusion} concludes with a perspective on applications to integrated CV quantum networks.

\begin{figure*}[b!]
    \centering
    \includegraphics[width=\textwidth]{Scheme.pdf}
    \caption{(a) Schematic setup of the entanglement distribution protocol where a source 
    located midway between Alice and Bob, separated by a distance $L$, emits an 
    entangled state $\hat{\rho}$ that is shared between them.
 (b) Main building blocks of the protocol, including the trimer-based state-generation 
module, the purification stage, and the photon-number–based teleportation 
module.
}
    \label{fig:scheme}
    \hrulefill
\end{figure*}

\section{Overview of the scheme}\label{sec:overview}

In this section, we present a high-level overview of the complete entanglement-distribution and teleportation protocol, as illustrated in Fig.~\ref{fig:scheme}. Our goal is to provide a unified description of how the non-Gaussian resource is generated on-chip, distributed to two remote parties through lossy channels, purified locally, and subsequently employed for photon-number–based teleportation. Detailed analyses of each component—generation, transmission, purification, and teleportation—are provided in Secs. \ref{sec:state_generation}–\ref{sec:teleportation}.

Our protocol begins with the preparation of a non-Gaussian entangled resource at a central station, after which the two output modes are distributed to Alice and Bob through independent optical channels. The full process unfolds through four sequential operations. First, a TMSVS is injected into the two outermost waveguides of an integrated waveguide trimer, labeled $a$ and $c$, while the central waveguide is initially prepared in the vacuum state. The three coupled waveguides then allow the TMSVS modes to interfere symmetrically through evanescent coupling.
This trimer configuration can be implemented using current integrated photonic technologies, such as femtosecond-laser-written waveguides~\cite{Szameit_2010, Perez-Leija2018}. The waveguide trimer has been studied as a versatile architecture for the generation and manipulation of multipartite quantum states, including multimode 
$W$ states~\cite{Perez-Leija:PRA:2013}, three-mode squeezed states~\cite{Rojas:PRA:2019}, and high-dimensional path-entangled states~\cite{Luo:PRA:2019}.
Mathematically, the TMSVS is defined as $\ket{\mathrm{TMSVS}} = \sqrt{1-|r|^{2}} \sum_{n=0}^{\infty} r^{n} \ket{n}_{a}\ket{n}_{c}$, where $r=\tanh{\xi}$, with $\xi$ being the parameter determining the squeezing strength (so that $|r|<1$), and modes $a$ and $c$ contain perfectly correlated photon numbers.
Conditional detection of photons in the central waveguide heralds the creation of quasi-ECS across the two outer modes. The characteristics of the generated state depend on the squeezing amplitude and the light-propagation distance inside the trimer (denoted $z$), which determines the interference pattern among the three waveguides. In our analysis we focus on the experimentally favorable regime in which a single photon is subtracted and the trimer is operated at one of its symmetric propagation points; however, the conceptual role of the generation stage does not rely on the specific numerical value of $z$.

After leaving the trimer, modes $a$ and $c$ propagate to Bob and Alice through optical channels characterized by a transmissivity $\eta$. Loss and decoherence introduce mixedness and reduce the fidelity of the quasi-ECS, representing the dominant source of degradation in the protocol.

Next, to counteract loss-induced degradation, each transmitted mode interferes with an ancillary single photon on a low-transmissivity directional coupler, as shown in Fig.~\ref{fig:scheme}(b). Heralding a single photon in each auxiliary port implements a single-photon catalysis operation, which selectively enhances the non-Gaussian structure of the resource and improves fidelity in relevant channel regimes, although it need not increase purity at every operating point. The resulting purified state serves as the effective shared entanglement resource.

The final operation is the teleportation step, implemented using the rightmost module in Fig.~\ref{fig:scheme}(b). 
Alice mixes her share of the purified quasi-ECS resource with the unknown input state $\ket{\psi_{0}}$ on a balanced beam splitter and performs photon-number-resolving detection on the two output ports. 
Each measurement outcome $(x,y)$ projects the overall state onto a corresponding conditional state on Bob’s side, denoted by $\hat{\rho}^{(x,y)}$. 
In the teleportation protocol of van Enk and Hirota~\cite{vanEnk2001}, a successful event corresponds to detecting an odd number of photons in one port while the other remains in vacuum, which ideally reproduces the input state $\ket{\psi_{0}}$ on Bob’s side. 
For practical implementation, we restrict attention to the lowest-order odd-photon outcomes, $(x,y)=(1,0)$ or $(0,1)$, corresponding to a single photon being heralded in one of Alice’s output ports while the other remains in vacuum. Alice communicates which outcome occurred, and Bob applies the corresponding phase correction specified in Sec.~\ref{sec:teleportation}.
The teleportation fidelity is defined as
\begin{equation}\label{eq:teleported_fidelity}
    F^{(i)}_{j} = \bra{\psi_{0}} \, \hat{\rho}^{(x,y)} \, \ket{\psi_{0}},
\end{equation}
where $\hat{\rho}^{(x,y)}$ denotes the normalized, phase-corrected conditional output state on Bob\'s side under a successful odd-photon detection event. 
Throughout the paper, we use the notation $F^{(i)}_{j}$, where the superscript $i$ labels the input state being teleported and the subscript $j$ identifies the entangled resource employed.

\begin{figure}[t!]
    \centering
    \includegraphics[width=0.55\textwidth]{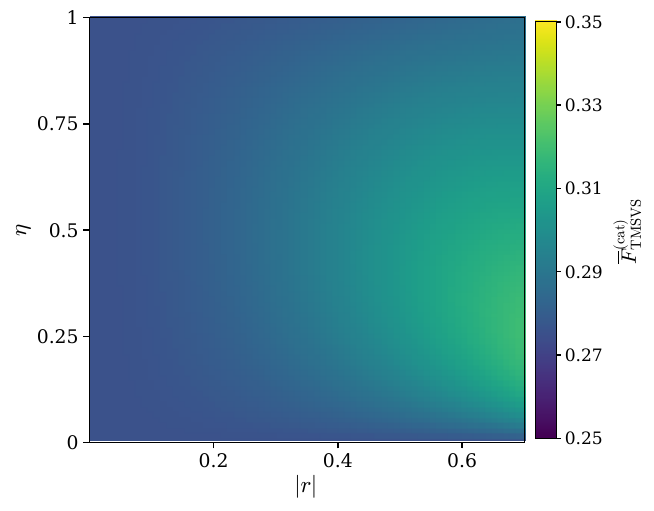}
    \caption{Average teleportation fidelity $F^{(\mathrm{cat})}_{\mathrm{TMSVS}}$ for teleporting a cat state using TMSVS as the entangled resource, plotted as a function of the channel transmissivity $\eta$ and the squeezing amplitude $|r|$, for $\beta=0.8$. The fidelity is the heralding-weighted average over the six orthogonal logical cat-qubit inputs and the two accepted photon-counting outcomes.}  
    \label{fig:Teleporation_TMSVS}
    \hrulefill
\end{figure}

Before turning to the detailed analysis of these building blocks, it is useful to address two natural questions that arise when introducing this protocol. The first question is: why not simply use a TMSVS, which is experimentally easier to generate? This is a reasonable possibility to examine. However, as shown in Fig.~\ref{fig:Teleporation_TMSVS}, the TMSVS remains below the standard qubit-teleportation benchmark of $2/3$ within the squeezing and channel-transmissivity ranges examined for $\beta=0.8$, under the specified photon-number measurement and correction procedure. In our analysis, we consider squeezing amplitudes up to $|r| = 0.7$, corresponding to experimentally accessible squeezing levels achievable in many present-day CV optical platforms~\cite{Magana-Loaiza:NQI:2019}. These results indicate that Gaussian entanglement alone is insufficient to support high-fidelity teleportation of non-Gaussian cat states within the photon-number-based operational framework tailored to coherent-state encodings, thereby motivating the use of non-Gaussian quasi-ECS resources. We emphasize that the comparison shown in Fig.~\ref{fig:Teleporation_TMSVS} is performed within the same van Enk--Hirota scheme employed throughout this work, allowing a direct assessment of how Gaussian and non-Gaussian entanglement perform under identical photon-number measurement conditions. In the ideal limit, the van Enk--Hirota protocol enables unit-fidelity teleportation of cat states when perfect odd entangled coherent states are used as resources~\cite{vanEnk2001}. This highlights that the protocol is naturally suited to coherent-state encodings and further motivates the use of quasi-ECS resources in the present setting. The comparison is specific to the chosen photon-counting measurement and feed-forward operations; it is not an optimization over all teleportation protocols using TMSVS resources. 
For completeness, we additionally consider later the standard Braunstein--Kimble (BK) teleportation protocol based on homodyne detection as a benchmark Gaussian teleportation scheme~\cite{Braunstein1998}.

\begin{figure}[t!]
    \centering
    \includegraphics[width=0.55\textwidth]{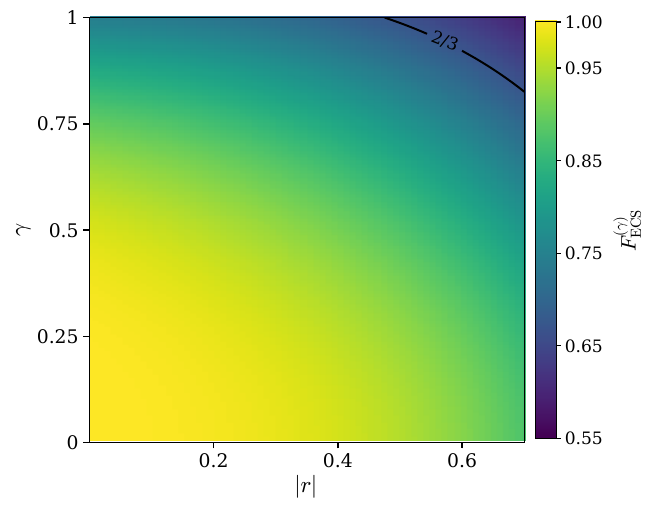}
    \caption{
    Average teleportation fidelity $F^{(\gamma)}_{\mathrm{ECS}}$ for teleporting a coherent state $\ket{\gamma}$ using unpurified quasi-ECS entangled resources as a function of the coherent-state amplitude $\gamma$ and the squeezing amplitude $|r|$, for $z = 1.25$, $N = 1$, and $\eta=1$. 
    The black contour line corresponds to $F^{(\gamma)}_{\mathrm{ECS}} = 2/3$, shown here as a reference level rather than a classical benchmark for the coherent-state input without a specified prior.}
    \label{fig:coherent_teleportation_fidelity}
    \hrulefill
\end{figure}

A second natural question arises from existing literature: before using the trimer-generated quasi-ECS resource to teleport a non-Gaussian cat-qubit, does it perform correctly in the standard task of teleporting a coherent state? 
This question is well motivated, since coherent-state teleportation has been extensively analyzed in the CV literature and demonstrated in several experimental and theoretical works~\cite{Furusawa1998, Zhao2023, Opatrny:PRA:2000, Zhang:PRA:2010}. It is therefore reasonable to verify that the trimer output behaves consistently with these established results. To this end, we evaluate the teleportation fidelity of an unknown coherent state using the quasi-ECS resource obtained via single-photon subtraction at a symmetric propagation point of the trimer. As shown in Fig.~\ref{fig:coherent_teleportation_fidelity}, the fidelity in the ideal lossless case exceeds the $2/3$ reference level across a broad range of parameters and approaches unity for suitable squeezing strengths. This provides a coherent-state performance check for the specified conditional protocol, before moving to the more demanding non-Gaussian regime.

Having established the teleportation capabilities of the resource, it is useful to place the underlying state-generation strategy in the context of related approaches. ECS have previously been generated by first producing cat states via photon subtraction from a single-mode squeezed vacuum and subsequently interfering them on a balanced beam splitter~\cite{Asavanant:17}. In addition, photon-subtracted TMSVSs have been investigated as non-Gaussian entanglement resources for cat-state teleportation~\cite{Seshadreesan_2015}. By contrast, the architecture considered here starts from a TMSVS and employs a symmetric waveguide trimer to generate quasi-ECS directly through a single photon-subtraction event, avoiding the sequential cat-state generation, beam-splitter interference, and coincidence-heralding stages used in earlier approaches.

\section{Generation of quasi-ECS}\label{sec:state_generation}

To generate the non-Gaussian entangled resources used in this work, we start with a TMSVS, in which the two modes exhibit perfectly correlated photon numbers. This state is injected into an integrated waveguide trimer, where the three waveguides are linearly coupled via evanescent interactions, enabling a symmetric mixing of the input modes through the central waveguide.
A key feature of this geometry is that photon subtraction from the central waveguide does not reveal which of the two input modes contributed the photon, because both modes couple identically to the middle channel. This erasure of “which-path” information is essential: detecting $N$ photons in the central waveguide projects modes $a$ and $c$ into a parity-defined superposition of photon-number components. Odd $N$ produces odd-parity states and even $N$ produces even-parity states. Indeed, the input TMSVS has even total photon-number parity and the trimer conserves total photon number; conditioning on $N=1$ leaves the two outer modes in an exactly odd-total-parity pure state prior to transmission loss.
In the ideal limit, these projected states approximate the ECS introduced in Section~\ref{sec:Introduction}. In this work we focus on odd quasi-ECS, as they offer stronger effective entanglement in the two-dimensional logical subspace relevant for photon-number–based teleportation~\cite{Park:PRA:2010, vanEnk2001}.

The photon-subtracted state $\hat{\rho}_{\rm sub}$ is obtained by conditioning on the detection of $N$ photons in the central waveguide. Its properties depend on the squeezing amplitude $|r|$ and the propagation distance $z$ within the trimer. Physically, $z$ controls the interference pattern generated by evanescent coupling among the three waveguides and therefore governs the properties of the parity-projected quasi-ECS produced by central-waveguide photon subtraction. The full analytical expression for $\hat{\rho}_{\rm sub}$ is provided in Appendix~\ref{sec:appendix_a}.

\begin{figure}[b!]
    \centering
    \includegraphics[width=1\textwidth,height=0.31\textwidth]{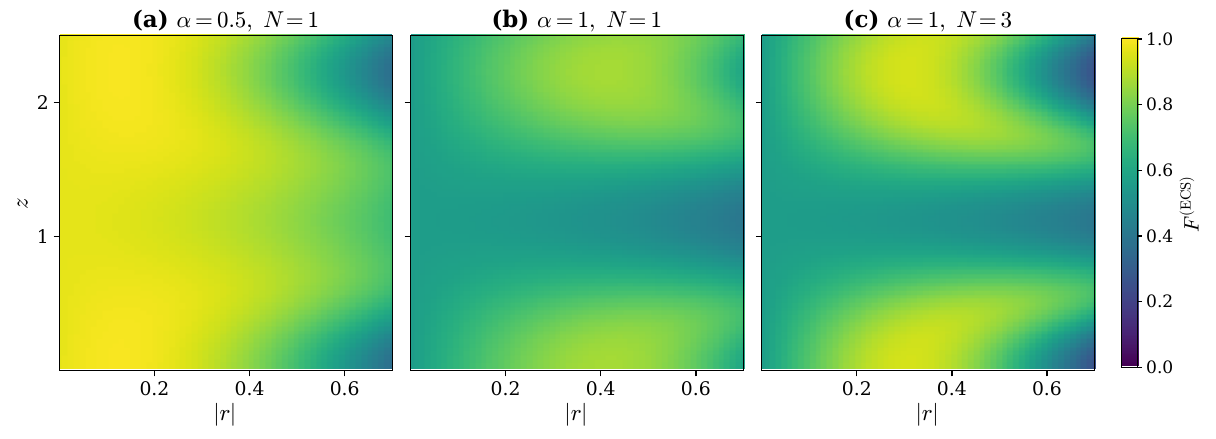}
    \caption{Fidelity of the quasi-ECS versus the propagation distance $z$ and squeezing amplitude $|r|$ for different values of the ECS amplitude $\alpha$ and the number of subtracted photons $N$.}
    \label{fig:ECS_fidelity}
    \hrulefill
\end{figure}

To assess how close the photon-subtracted states are to ideal odd ECS, we compute the fidelity $F(\rm{ECS}) = \langle \psi_{\mathrm{ECS}}^{-} | \hat{\rho}_{\rm sub} | \psi_{\mathrm{ECS}}^{-} \rangle$ and analyze its dependence on the trimer propagation distance $z$, the squeezing amplitude $|r|$, the photon-subtraction number $N$, and the target ECS amplitude $\alpha$.
The analytical expression of $F$(ECS) has been derived in Appendix~\ref{sec:appendix_a}. This expression makes explicit how the fidelity depends on the propagation distance through the trimer via the interference parameter $\Theta$, as well as on the squeezing amplitude.
Fig.~\ref{fig:ECS_fidelity}(a) shows $F(\rm{ECS})$ for $\alpha=0.5$ and $N=1$, where the maximum is nearly unity. Increasing $|r|$ reduces the fidelity, and for larger ECS amplitudes ($\alpha=1$) with the same number of subtracted photons, the fidelity is lower (Fig.~\ref{fig:ECS_fidelity}(b)). Higher fidelity for larger $\alpha$ can be achieved by increasing $N$ (e.g., $N=3$ in Fig.~\ref{fig:ECS_fidelity}(c)). The plots exhibit symmetry with respect to $z$, reflecting the periodic intensity distribution among the three waveguides. For the remainder of this work, we set $z=1.25$ corresponding to this symmetric configuration and focus on $N=1$, because single-photon heralding is the simpler generation event considered in our principal teleportation analysis. The generation probability depends on $r$ and $z$, and it is included explicitly when reporting end-to-end success probabilities.

At this stage, it is useful to reiterate that our theoretical framework relies on the assumption of ideal waveguide propagation, namely, equal propagation constants and coupling coefficients that remain uniform along the propagation direction. This assumption is well justified in light of the capabilities of modern fabrication platforms, particularly femtosecond-laser-written waveguide arrays~\cite{Szameit_2010}. Such systems exhibit propagation losses as low as $0.3\,\mathrm{dB/cm}$ and enable precise control over both propagation constants and inter-waveguide coupling strengths~\cite{Weimann2016, Selim:Science:2025}. Moreover, recent photon-number-resolving detectors have demonstrated quantum efficiencies exceeding $99\%$~\cite{Li:LSA:2025}. Therefore, imperfections associated with waveguide propagation, state generation, and photon detection are expected to be small in suitably engineered experimental platforms. Nevertheless, their quantitative influence is system dependent and has not been included in the present theoretical model.

In the context of quantum communication between distant parties, however, the dominant source of attenuation arises from transmission through the communication channel used to distribute entanglement. This effect is typically characterized by the channel transmissivity $\eta$, which decreases exponentially with transmission distance. In the remainder of this work, we therefore focus exclusively on the distance-dependent channel transmissivity as the primary loss mechanism.

\section{Channel loss and purification}\label{sec:loss_purification}

\begin{figure}[t!]
    \centering
    \includegraphics[width=0.95\textwidth,height=0.74\textwidth]{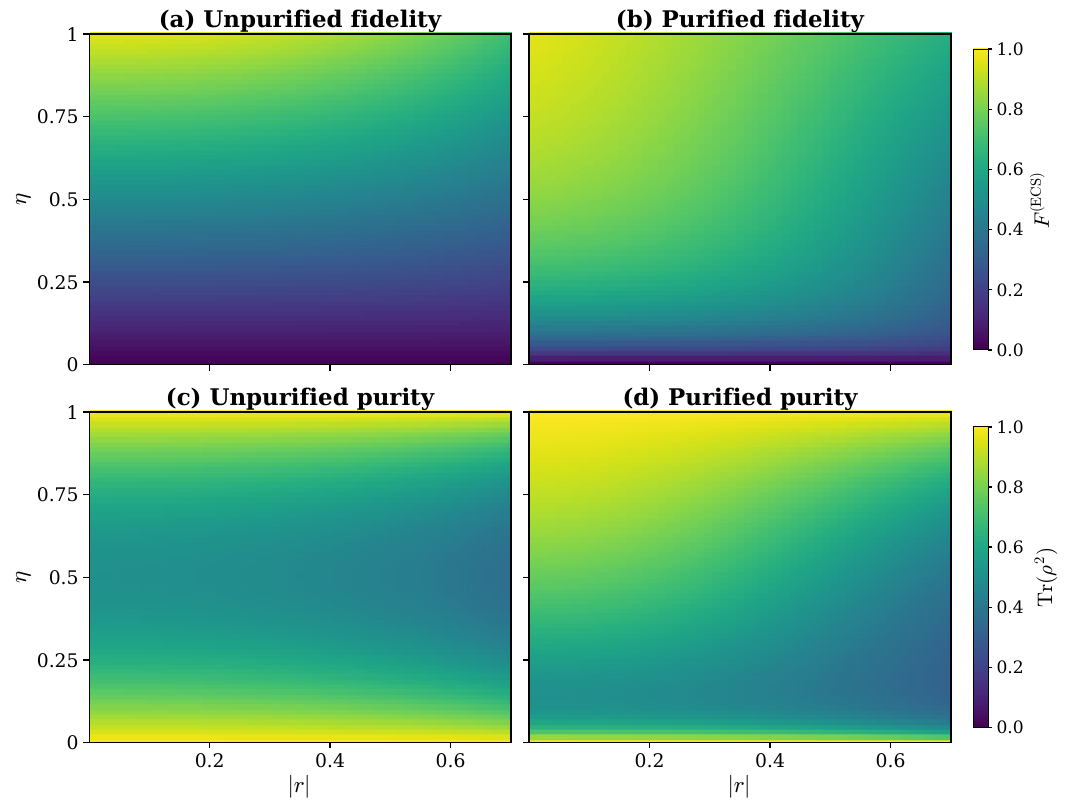}
    \caption{
    Fidelity and purity of the quasi-ECS as functions of the channel transmissivity $\eta$ and squeezing parameter $|r|$, for $z = 1.25$ and $N = 1$. 
    The fidelity plots (top row) correspond to $\alpha = 0.5$: (a) before and (b) after purification using single-photon catalysis implemented via a directional coupler with transmission coefficient $T = 0.1$. 
    The bottom row presents the corresponding purity (c) before and (d) after purification.
    }
    \label{fig:ECS_purified_fidelity}
    \hrulefill
\end{figure}

After leaving the trimer, the two modes $a$ and $c$ propagate to Bob and Alice through independent channels of transmissivity $\eta$. Photon loss during transmission transforms the pure quasi-ECS into a mixed state $\hat{\rho}_{\rm lossy}$, whose analytical form is given in Appendix~\ref{sec:appendix_b}. To quantify this degradation, we compute the fidelity $F'(\rm{ECS}) = \langle \psi_{\mathrm{ECS}}^{-} | \hat{\rho}_{\mathrm{lossy}} | \psi_{\mathrm{ECS}}^{-} \rangle$. Fig.~\ref{fig:ECS_purified_fidelity}(a) shows $F'(\rm{ECS})$ as a function of $\eta$ and the squeezing parameter $|r|$ for $\alpha=0.5$ and $N=1$. As expected, increasing channel loss reduces the fidelity, whereas the dependence on $|r|$ is comparatively mild within the considered range.

To counteract the degradation induced by channel loss, we introduce a purification protocol based on single-photon catalysis, applied locally by Alice and Bob on the received modes. As depicted in Fig.~\ref{fig:scheme}(b), the output modes $a$ and $c$ of the trimer—after propagating through the lossy channels—each interfere with an injected single photon in ancillary modes $a'$ and $c'$ via low-transmissivity directional couplers. Heralding a single-photon detection in both ancillary ports conditionally purifies the shared quasi-ECS resource.

We now examine how purification affects the distributed state. Using a fixed directional-coupler transmissivity of $T=0.1$ for the representative operating regime considered here, we evaluate the fidelity of the purified state $\hat{\rho}_{\rm purified}$ as a function of the channel transmissivity $\eta$ and the squeezing parameter $|r|$. As shown in Fig.~\ref{fig:ECS_purified_fidelity}(b), the purification step substantially increases the fidelity compared to the unpurified case in Fig.~\ref{fig:ECS_purified_fidelity}(a). To further characterize the quality of the purified resource, we also compute its purity, defined as $P = \mathrm{Tr}(\hat{\rho}_{\mathrm{purified}}^{2})$. Figs.~\ref{fig:ECS_purified_fidelity}(c) and \ref{fig:ECS_purified_fidelity}(d) show the purity before and after purification, respectively. As $\eta \to 0$, the purity approaches unity since the state approximates a vacuum. Similarly, for $\eta \to 1$, the state remains pure, $P = 1$. The purity may increase in parts of the intermediate-loss regime but does not improve at every operating point; the resource-fidelity improvement should therefore be distinguished from a universal increase in purity.

We also compute the success probability of the purification step, defined as the probability of heralding a single photon in both ancillary ports during the single-photon–catalysis events on modes $a$ and $c$. The success probability is defined as
\begin{align}\label{eq:purification_probability}
    \mathcal{P}_{\mathrm{purified}} = \mathrm{Tr}(\widetilde{\rho}_{\mathrm{purified}}),
\end{align}
where the unnormalized purified state is
\begin{align}\label{eq:purified_state}
\widetilde{\rho}_{\rm purified} =
\leftindex{_{a'_{\rm out}}}{\langle 1|} \, 
\leftindex{_{c'_{\rm out}}}{\langle 1|} \, 
\Big[ \hat{U}_{\rm tot} \,
\Big( \hat{\rho}_{\rm lossy} \otimes 
\ket{1}\bra{1}_{a'_{\rm in}} \otimes 
\ket{1}\bra{1}_{c'_{\rm in}} \Big) 
\hat{U}_{\rm tot}^\dagger \Big] 
\ket{1}_{a'_{\rm out}} \, 
\ket{1}_{c'_{\rm out}},
\end{align}
with the total unitary
$\hat{U}_{\rm tot} = \hat{U}_{a,a'} \otimes \hat{U}_{c,c'}$,
where $\hat{U}_{a,a'}$ and $\hat{U}_{c,c'}$ are the directional-coupler unitaries acting on modes $(a,a')$ and $(c,c')$, respectively. The ancillary input ports of these couplers are labeled $a'_{\mathrm{in}}$ and $c'_{\mathrm{in}}$, and the corresponding output ports are denoted by $a'_{\mathrm{out}}$ and $c'_{\mathrm{out}}$.

\begin{figure}[t!]
    \centering
    \includegraphics[width=0.55\textwidth]{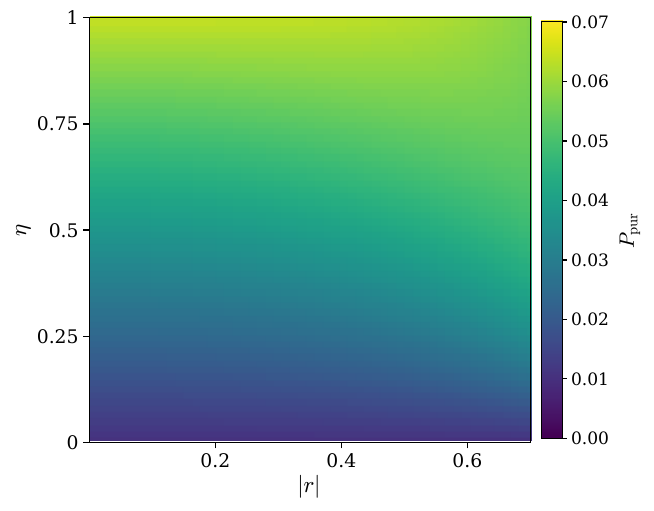}
    \caption{Probability of successful purification for quasi-ECS as a function of the channel transmissivity $\eta$ and the squeezing amplitude $|r|$, for $z = 1.25$, $N = 1$, and a directional coupler transmission coefficient of $T = 0.1$.  
    }
    \label{fig:purified_probability}
    \hrulefill
\end{figure}

Fig.~\ref{fig:purified_probability} shows $\mathcal{P}_{\rm purified}$ as a function of the channel transmissivity $\eta$ and the squeezing amplitude $|r|$, while keeping other parameters the same as in the purity calculations. As expected, the success probability decreases with increasing channel loss, since fewer photons survive the lossy channel for successful heralding. 


\section{Teleportation protocol and performance}\label{sec:teleportation}

With a purified quasi-ECS shared between Alice and Bob, we now turn to the teleportation stage. In this section, we evaluate the performance of the non-Gaussian teleportation protocol and focus on the teleportation of Schrödinger-cat states. The input state is taken to be
$\ket{\psi_{0}} = (\epsilon_{+}\ket{\beta} + \epsilon_{-}\ket{-\beta})/\sqrt{\mathcal{N}_{0}}$, where $\beta$ 
denotes the amplitude of the cat state to be teleported, and $\mathcal{N}_{0} = |\epsilon_{+}|^{2} + |\epsilon_{-}|^{2} + 2 c_{\beta}\,\mathrm{Re}[\epsilon_{+}^{*}\epsilon_{-}]$ is the normalization factor. 
To perform the teleportation, we adopt the ECS-based protocol introduced by van Enk and Hirota~\cite{vanEnk2001}, which differs from the standard CV scheme based on homodyne detection. Instead of quadrature measurements, the protocol uses photon-number resolution. Alice mixes the state to be teleported with her half of the entangled channel—prepared as an odd ECS—on a balanced beam splitter and performs photon-number detection on the two output ports. Successful teleportation is achieved when one of the detectors registers an odd photon number while the other registers vacuum, in which case Bob obtains a conditional state whose recovery requires the appropriate measurement-dependent correction.

In our scheme, the purified quasi-ECS serves as the shared entangled resource for teleportation, with one mode held by Bob and the other by Alice (see Fig.~\ref{fig:scheme}(b)). The unknown state $\ket{\psi_{0}}$ is combined with Alice’s resource mode on a balanced coupler, and each measurement outcome $(x,y)$ projects Bob’s mode onto the conditional state $\hat{\rho}^{(x,y)}$. The teleportation fidelity for the lowest-order odd-photon events, $(x,y)=(1,0)$ or $(0,1)$, is given by $F^{\rm{(cat)}}_{\mathrm{ECS}}$ as defined in Eq.~\eqref{eq:teleported_fidelity}.
With the balanced beam-splitter convention of Appendix~\ref{sec:appendix_d}, Bob applies a phase correction determined by Alice's reported outcome,
\begin{align}\label{eq:bob_feedforward}
\hat C_{10}=\exp\!\left(-\frac{i\pi}{2}\hat n_a\right),\qquad
\hat C_{01}=\exp\!\left(+\frac{i\pi}{2}\hat n_a\right),
\end{align}
where $\hat n_a$ denotes Bob's photon-number operator. These corrections act on Bob's output mode; the original optical arrangement in Fig.~\ref{fig:scheme} already indicates the required classical feed-forward. No additional phase shifter is placed in Alice's input arm.

\begin{figure}[t!]
    \centering
    \includegraphics[width=0.9\textwidth,height=0.4\textwidth]{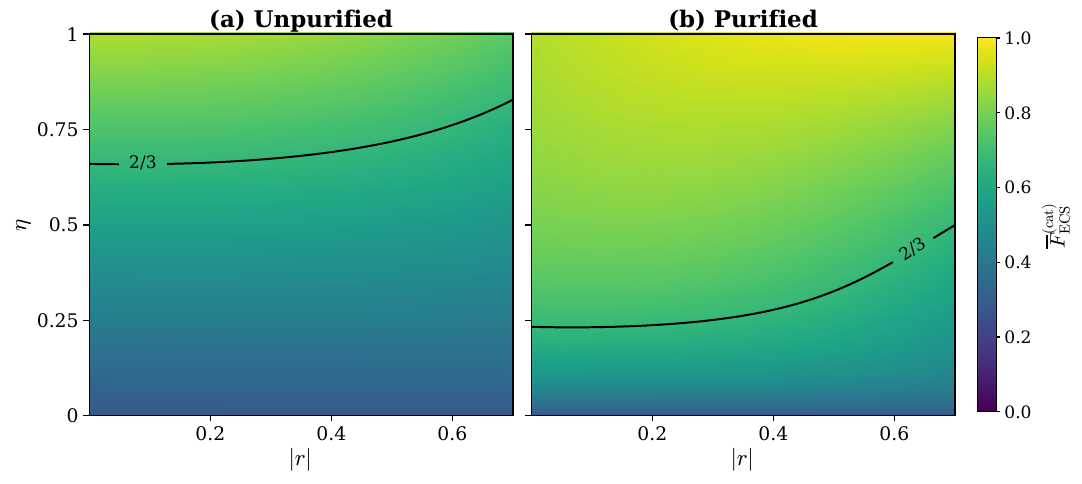}
    \caption{
    Average teleportation fidelity $F^{\mathrm{(cat)}}_{\mathrm{ECS}}$ for teleporting a cat state with amplitude $\beta = 0.8$ as a function of the channel transmissivity $\eta$ and the squeezing amplitude $|r|$, for $z = 1.25$, $N = 1$, and $T = 0.1$:
    (a) before and (b) after purification using single-photon catalysis implemented via directional couplers. The fidelity is success-probability weighted over the six orthogonal logical cat-qubit states and both accepted detector outcomes, after Bob's phase correction.
    The black contour line corresponds to $F^{\mathrm{(cat)}}_{\mathrm{ECS}} = 2/3$, which represents the classical fidelity limit for teleporting a qubit using a purely classical strategy. Regions above this contour indicate genuinely quantum teleportation performance.
    }
    \label{fig:Teleportation_fidelity}
    \hrulefill
\end{figure}

%
To comprehensively characterize the teleportation performance, we compute the average fidelity over six representative cat states corresponding to different choices of $(\epsilon_{+}, \epsilon_{-})$, analogous to the six-state sampling used to span the Bloch sphere in the DV qubit case~\cite{Dong:npj:2021}. In the DV case, the six states are given by
\begin{align}
\bigg\{
\ket{0},\,
\ket{1},\,
\frac{\ket{0}\pm\ket{1}}{\sqrt{2}},\,
\frac{\ket{0}\pm i\ket{1}}{\sqrt{2}}
\bigg\}.
\end{align}
In our teleportation protocol, we adopt a cat-state encoding in which the logical computational basis is defined by the normalized even- and odd-parity cat states,
\begin{align}
\ket{0}_L \equiv \ket{C_\beta^+}
&=
\frac{\ket{\beta}+\ket{-\beta}}
{\sqrt{2(1+e^{-2|\beta|^2})}},\\
\ket{1}_L \equiv \ket{C_\beta^-}
&=
\frac{\ket{\beta}-\ket{-\beta}}
{\sqrt{2(1-e^{-2|\beta|^2})}}.
\end{align}
Since the even- and odd-parity cat states are orthonormal, the six logical input states corresponding to the DV states are therefore
\begin{align}
\bigg\{
&\ket{C_\beta^+},\,
\ket{C_\beta^-},\,
\frac{\ket{C_\beta^+}\pm \ket{C_\beta^-}}{\sqrt{2}},\,
\frac{\ket{C_\beta^+} \pm i\ket{C_\beta^-}}{\sqrt{2}}
\bigg\}.
\end{align}
For each equally likely logical input $\ket{\psi_k}$ ($k=1,\ldots,6$) and accepted photon-counting outcome $m\in\mathcal M=\{(1,0),(0,1)\}$, let $\widetilde\rho_{k,m}^{\rm corr}$ be Bob's unnormalized output after the corresponding phase correction. Its trace is the conditional outcome probability $p_{k,m}=\mathrm{Tr}(\widetilde\rho_{k,m}^{\rm corr})$, and $F_{k,m}=\bra{\psi_k}\widetilde\rho_{k,m}^{\rm corr}\ket{\psi_k}/p_{k,m}$ is its normalized fidelity. The success-probability-weighted conditional average is
\begin{align}\label{eq:weighted_six_average}
\overline F_{\rm succ}=\frac{\displaystyle\sum_{k=1}^{6}\sum_{m\in\mathcal M}p_{k,m}F_{k,m}}{\displaystyle\sum_{k=1}^{6}\sum_{m\in\mathcal M}p_{k,m}}.
\end{align}
Unlike a simple average of the six individually normalized fidelities, this expression accounts for input-dependent heralding probabilities. We use this measure for the postselected logical cat-qubit results in Figs.~\ref{fig:Teleporation_TMSVS} and \ref{fig:Teleportation_fidelity}.
We then analyze the average teleportation fidelity $F^{\rm{(cat)}}_{\rm{ECS}}$ for teleporting a cat state with amplitude $\beta = 0.8$ as a function of the squeezing amplitude $|r|$ and the channel transmissivity $\eta$, while keeping all other parameters identical to those used in the purification analysis. The results for the unpurified and purified quasi-ECS resources are shown in Fig.~\ref{fig:Teleportation_fidelity}(a) and (b), respectively. 
The black contour line marks the benchmark fidelity value $F^{\rm (cat)}_{\rm ECS}=2/3$.
As shown in Fig.~\ref{fig:Teleportation_fidelity}, our protocol achieves fidelities surpassing this benchmark over a broad parameter range, demonstrating the viability of quasi-ECS as entangled resources for CV quantum teleportation. 
Moreover, the purification protocol substantially extends the operational range in $\eta$ (Fig.~\ref{fig:Teleportation_fidelity}(b)), allowing high-fidelity teleportation over longer effective channel distances compared to the unpurified case (Fig.~\ref{fig:Teleportation_fidelity}(a)).

The overall probability of successful teleportation includes the initial trimer heralding event. Let $\mathcal P_{\rm gen}=\mathcal N_{\rm sub}$ be the probability of generating the quasi-ECS, and let $p_{k,m}^{\rm(unpur)}$ and $p_{k,m}^{\rm(pur)}$ denote the probabilities of the accepted teleportation outcomes conditioned on the normalized unpurified and purified resources, respectively. For uniformly sampled logical inputs we obtain
\begin{align}\label{eq:unpurified_teleportation}
\mathcal P_{\rm tot}^{\rm(unpur)} = \frac{\mathcal P_{\rm gen}}{6}\sum_{k=1}^{6}\sum_{m\in\mathcal M}p_{k,m}^{\rm(unpur)},\nonumber\\
\mathcal P_{\rm tot}^{\rm(pur)} = \frac{\mathcal P_{\rm gen}\mathcal P_{\rm purified}}{6}\sum_{k=1}^{6}\sum_{m\in\mathcal M}p_{k,m}^{\rm(pur)}.
\end{align}
Here $\mathcal P_{\rm purified}$ is the catalysis heralding probability conditional on successful resource generation and transmission, as defined in Eq.~\eqref{eq:purification_probability}. The purified teleportation outcome probabilities are calculated with the purified state; they are not generally equal to those for the unpurified resource.

\begin{figure}[H]
    \centering
    \includegraphics[width=0.9\textwidth]{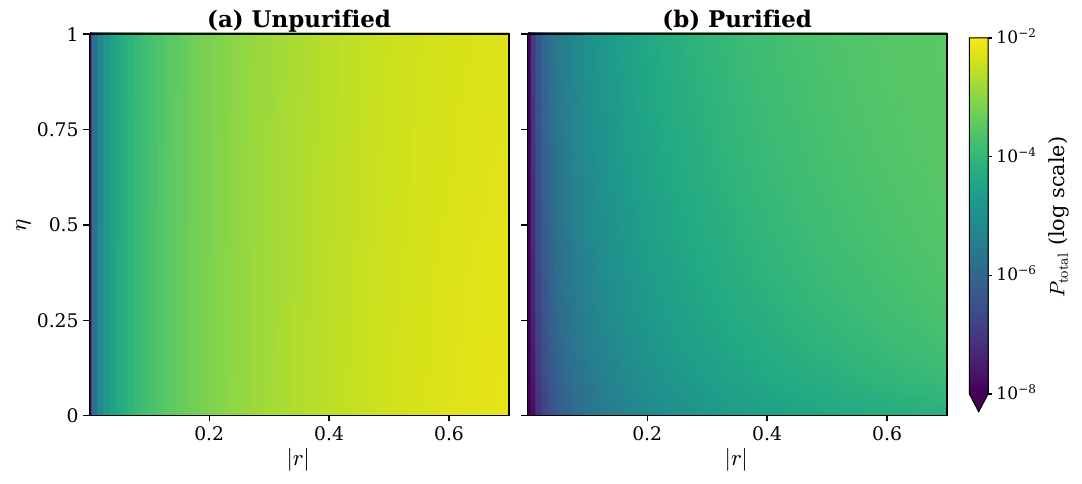}
    \caption{Success probability $\mathcal{P}_{\mathrm{tel}}$ of the teleportation protocol as a function of the channel transmissivity $\eta$ and squeezing amplitude $|r|$: 
    (a) unpurified and (b) purified quasi-ECS entangled resources. 
    All other parameters are identical to those used for the fidelity plots in Fig.~\ref{fig:Teleportation_fidelity}. 
    Both panels show the end-to-end probability, including trimer generation and both accepted detector outcomes; the purified panel also includes catalysis success. A common logarithmic colour scale is used to display the different probability ranges.}
    \label{fig:Teleportation_probability}
    \hrulefill
\end{figure}

Fig.~\ref{fig:Teleportation_probability} shows these end-to-end success probabilities as functions of $\eta$ and $|r|$ for (a) unpurified and (b) purified quasi-ECS resources. Both accepted photon-counting outcomes are included. Purification reduces the overall probability through its additional heralding requirement, even where it substantially improves the conditional fidelity.

The small amplitudes considered for the input cat state ($\beta=0.8$) and the quasi-ECS ($\alpha=0.5$) provide a useful perspective on the teleportation protocol. The input cat state closely approximates a single-rail qubit, while the quasi-ECS approximates a single photon delocalized between two modes. Accordingly, the teleportation task can be interpreted approximately as the teleportation of an unknown single-rail qubit through a nonlocal quantum channel, as discussed in Refs.~\cite{Lombardi:PRL:2002,Lee:PRA:2000}. This perspective also motivates exploring hybrid quantum channels that combine DV and CV encodings as a possible avenue for improving teleportation performance. Such an extension could be investigated in future work, particularly in terms of teleportation fidelity and transmission efficiency \cite{Jeong:QIP:2016,Ulanov:PRL:2017,Podoshvedov:LPL:2026}.


\Needspace{14\baselineskip}
To examine parity preservation directly, we evaluate Bob's photon-number parity $\hat\Pi_B=(-1)^{\hat n_a}$ for the two logical $Z$-basis inputs. The probability of correct output parity is $P_{+}=(1+\langle\hat\Pi_B\rangle_{C_+})/2$ for an even-cat input and $P_{-}=(1-\langle\hat\Pi_B\rangle_{C_-})/2$ for an odd-cat input. Table~\ref{tab:output_parity} summarizes the conditional results. Under loss, the unpurified output tends towards vacuum, making even parity appear favourable even when the teleportation fidelity is low. Purification improves odd-cat parity preservation in this regime, but does not increase the even-cat parity fraction at all operating points.
\begin{table}[H]
\centering
\caption{Probability of the correct output photon-number parity for an even-cat ($P_+$) or odd-cat ($P_-$) input, before and after purification. The cat amplitude is $\beta=0.8$, with $|r|=0.3$, $z=1.25$, $N=1$, and $T=0.1$. The two successful photon-counting outcomes are included.}
\label{tab:output_parity}
\small
\begin{tabular}{c c c c c}
\hline
$\eta$ & $P_+^{\rm unpur}$ & $P_+^{\rm pur}$ & $P_-^{\rm unpur}$ & $P_-^{\rm pur}$\\
\hline
1.0 & 1.000 & 1.000 & 1.000 & 1.000\\
0.8 & 0.990 & 0.946 & 0.664 & 0.921\\
0.5 & 0.985 & 0.914 & 0.337 & 0.759\\
0.2 & 0.990 & 0.939 & 0.117 & 0.458\\
\hline
\end{tabular}
\end{table}

\begin{figure}[t!]
\centering
\includegraphics[width=0.95\textwidth]{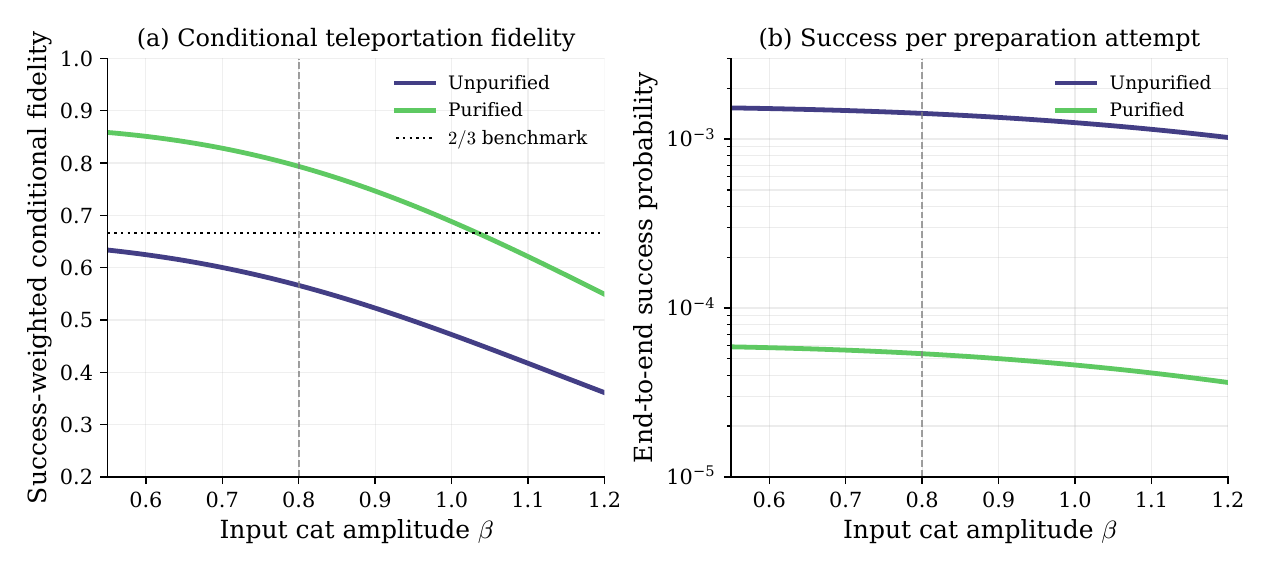}
\caption{Dependence of the teleportation performance on the input cat-state amplitude $\beta$ for $|r|=0.3$, $\eta=0.5$, $z=1.25$, $N=1$, and $T=0.1$. (a) Success-probability-weighted conditional average fidelity over the six orthogonal cat-qubit inputs, for unpurified and purified quasi-ECS resources; the dotted line marks the classical-qubit threshold $2/3$. (b) Corresponding end-to-end success probabilities, including resource generation, the accepted teleportation outcomes, and purification where applicable.}
\label{fig:cat_amplitude_scan}
\hrulefill
\end{figure}

\Needspace{10\baselineskip}
We also examine the dependence on the cat amplitude $\beta$, keeping the resource and optical parameters fixed. Figure~\ref{fig:cat_amplitude_scan} shows the success-weighted conditional fidelity and the end-to-end heralding probability at $|r|=0.3$ and $\eta=0.5$. The purified fidelity decreases from approximately $0.859$ at $\beta=0.55$ to $0.794$ at $\beta=0.8$ and $0.549$ at $\beta=1.2$. This loss of fidelity at larger amplitudes reflects increasing mismatch between the input cat's photon-number distribution and the resource and one-photon teleportation measurement.

Although the present analysis focuses on relatively small cat amplitudes, extensions to larger amplitudes may be possible through further engineering of the entangled resource. One approach is to optimise the squeezing, trimer propagation length and catalysis parameters to obtain a photon-number distribution better matched to the input cat states. More general photon-number-selective operations could provide additional control.

\begin{figure}[b!]
    \centering
    \includegraphics[width=0.55\textwidth]{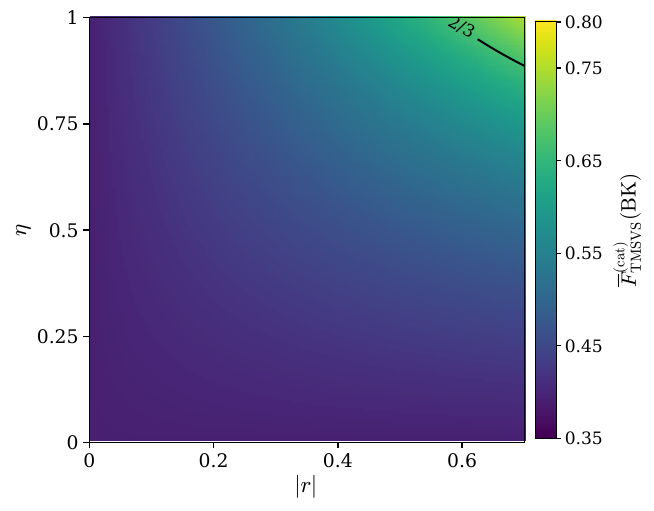}
    \caption{Average teleportation fidelity $F^{(\mathrm{cat})}_{\mathrm{TMSVS}}(\mathrm{BK})$ for teleporting a cat state using TMSVS as the entangled resource within the BK protocol, plotted as a function of the channel transmissivity $\eta$ and the squeezing amplitude $|r|$, for $\beta=0.8$. The black contour line corresponds to $F^{(\mathrm{cat})}_{\mathrm{TMSVS}}(\mathrm{BK}) = 2/3$, which represents the classical fidelity limit for teleporting a qubit using a purely classical strategy. The BK model uses unit-gain feed-forward and identical pure-loss channels on both resource modes.} Regions above this contour indicate genuinely quantum teleportation performance.
    \label{fig:BK_teleportation_fidelity}
    \hrulefill
\end{figure}

To further benchmark the teleportation performance beyond the photon-number--based protocol analyzed previously, and to compare with a readily available Gaussian entangled resource, we additionally consider the BK teleportation protocol~\cite{Braunstein1998}, which employs balanced homodyne detection together with a TMSVS as the entangled resource. In the BK protocol, Alice combines the input state with one mode of the shared TMSVS on a balanced beam splitter and performs joint quadrature measurements using homodyne detection. The measurement outcomes are then classically communicated to Bob, who applies the corresponding phase-space displacement operation to reconstruct the teleported state.
For a TMSVS resource with squeezing parameter $\xi$ whose two modes each undergo an independent pure-loss channel of transmissivity $\eta$, with unit-gain displacement feed-forward, the teleportation process can be modeled as a Gaussian random displacement channel acting on the input state $\hat{\rho}_{\mathrm{in}}$. The teleported output state is given by
\begin{align}
\hat{\rho}_{\mathrm{out}}
=
\int d^{2}\gamma 
P(\gamma)
\hat{D}(\gamma)
\hat{\rho}_{\mathrm{in}}
\hat{D}^{\dagger}(\gamma),
\end{align}
where $\hat{D}(\gamma)$ is the displacement operator and
\begin{align}
P(\gamma)
=
\frac{1}{\pi \sigma}
\exp\left(
-\frac{|\gamma|^{2}}{\sigma}
\right)
\end{align}
is the Gaussian noise distribution with variance
\begin{align}
\sigma=\eta e^{-2\xi}+(1-\eta).
\end{align}
Here, the first term represents the finite-squeezing noise after symmetric attenuation of the TMSVS modes, while the second term represents the vacuum noise introduced by the two pure-loss channels. In the ideal limit $\xi \rightarrow \infty$ and $\eta \rightarrow 1$, the added Gaussian noise vanishes and perfect teleportation is recovered.
The fidelity of the teleported unknown cat state is then computed using Eq.~\eqref{eq:teleported_fidelity}, where $\hat{\rho}_{\mathrm{out}}$ is the teleported output state.
For a direct comparison with the quasi-ECS results, we average the teleportation fidelity over the same six states introduced previously.
The resulting average fidelity is then evaluated as a function of the channel transmissivity $\eta$ and squeezing amplitude $|r|$ for $\beta = 0.8$, as shown in Fig.~\ref{fig:BK_teleportation_fidelity}. The black contour line denotes the benchmark fidelity value of $2/3$.
As shown in Fig.~\ref{fig:BK_teleportation_fidelity}, the BK protocol achieves fidelities above this benchmark in the strong-squeezing regime. However, such performance is sustained only up to approximately $\eta \sim 0.88$. For transmissivities below this value, corresponding to increased channel loss and longer transmission distances, the teleportation fidelity falls below the benchmark due to the rapid degradation of the nonclassical correlations in the TMSVS resource.
In terms of conditional teleportation fidelity, the photon-number-based van Enk--Hirota protocol employing the quasi-ECS resource remains above the $2/3$ benchmark over a broader range of channel transmissivities than the standard BK protocol considered here. This conditional-fidelity advantage must be considered alongside the reduced end-to-end heralding probability shown in Fig.~\ref{fig:Teleportation_probability}.

\section{Summary and Conclusions}\label{sec:conclusion}

In this work, we have developed and analyzed an integrated-photonics protocol for generating, distributing, purifying, and utilizing non-Gaussian entanglement suitable for CV quantum teleportation. Starting from a TMSVS, we have employed an integrated waveguide trimer to perform symmetric mode mixing such that heralded photon subtraction in the central waveguide produces non-Gaussian parity-projected states closely approximating odd ECS. This mechanism exploits the intrinsic mode symmetry of the trimer to erase which-path information, enabling chip-compatible generation of quasi-ECS entanglement without requiring strong optical nonlinearities or bulk interferometric stabilization.

We have then examined the degradation of these quasi-ECS under lossy channels using a full Fock-basis model. To counteract loss-induced decoherence, we have introduced a purification protocol based on single-photon catalysis applied locally by Alice and Bob. This purification step substantially enhances the fidelity of the shared resource in the operating regime investigated, although it does not improve purity for every transmissivity, while remaining compatible with realistic directional couplers and heralded single-photon sources. These improvements are accompanied by a reduction in the overall heralding probability, reflecting the tradeoff between resource quality and probabilistic implementation.

Using the purified quasi-ECS as resources, we have analyzed the photon-number-based teleportation protocol of van Enk and Hirota. We have shown that coherent-state teleportation achieves high fidelity, validating the resource in a well-understood setting. More importantly, we have demonstrated that the teleportation of cat states surpasses the benchmark fidelity value of $2/3$ over a wide parameter range. Furthermore, we find that purification enhances the teleportation performance by extending the range of channel transmissivities over which the success-probability-weighted conditional average fidelity exceeds the \(2/3\) benchmark. This demonstrates that the improvement in resource quality translates into a broader operational regime for high-fidelity quantum teleportation, although at the expense of a reduced overall success probability due to the additional heralding step.
By contrast, when the same photon-number-based protocol is implemented using a Gaussian TMSVS resource, the corresponding teleportation fidelity remains below this benchmark across the parameter range considered. This contrast highlights the importance of the entangled resource in determining teleportation performance. Although the ideal van Enk--Hirota protocol can achieve high fidelity in the multiphoton regime using suitable entangled coherent-state resources, the corresponding performance with a Gaussian TMSVS resource remains limited over the parameter range considered.
These results highlight the effectiveness of the integrated generation and purification of quasi-ECS entanglement within the photon-number–based teleportation architecture. For comparison, we have also investigated the BK teleportation protocol based on a Gaussian TMSVS resource and found that the quasi-ECS protocol maintains conditional teleportation fidelity above the $2/3$ benchmark over a broader range of channel transmissivities than this BK reference, albeit at the cost of a reduced end-to-end heralding probability.

More generally, the performance of cat-state teleportation depends on the correlations of the entanglement resource and on their compatibility with the chosen teleportation protocol. In the van Enk--Hirota protocol, the required photon-number detection outcomes, including the vacuum condition in one of Alice's output modes, play a central role in determining the conditional output state and the heralding probability \cite{vanEnk2001,Wang:PRA:2001}. The Braunstein--Kimble protocol, in contrast, relies on joint quadrature measurements and feed-forward operations \cite{Braunstein1998}. The present analysis is restricted to the low-amplitude cat-qubit regime considered here, with \(\beta=0.8\), and does not establish optimal teleportation performance for arbitrary cat states. For larger cat amplitudes, the van Enk--Hirota protocol is expected to become increasingly sensitive to matching between the coherent-state amplitudes of the input and entangled resource. In particular, a mismatch reduces the probability of the required event in which one output port is projected onto vacuum, while resource imperfections and loss may degrade the photon-number-parity structure of the conditional output state. The output-parity and amplitude-dependence calculations presented in Sec.~\ref{sec:teleportation} quantify these limitations for the specified low-amplitude regime; broader optimization of resource--protocol combinations remains an important direction for future work.

A detailed analysis of additional experimental imperfections is beyond the scope of this work, as their impact is highly system dependent~\cite{Furusawa1998, Zhao2023, Dong:npj:2021} and determined by the properties of the TMSVS source~\cite{Magana-Loaiza:NQI:2019, Sperling:PRL:2015}, waveguide parameters~\cite{Szameit_2010, Politi:Science:2008}, single-photon source performance~\cite{Meyer:RSI:2023}, and photon-number-resolving detector characteristics~\cite{Hadfield:23, Wang:AOM:2025}.

Our results highlight the potential of integrated photonics as a practical platform for non-Gaussian state engineering and CV quantum communication~\cite{Nishio2025multiplexedquantum, Laurent:PRX:2024}. The proposed approach can be extended to higher-amplitude ECS, multiplexed entanglement distribution, and fault-tolerant bosonic encodings relevant for quantum repeaters and distributed quantum computation~\cite{Li_2025, Li:AQT:2024}. As integrated sources, detectors, and nonlinear elements continue to advance, the methods demonstrated here offer a promising route toward scalable non-Gaussian quantum networks~\cite{Walschaers:PRX:2021, Walschaers_2023}.

\ack{This work was supported by the Moonshot R \& D Program Grant JPMJMS226C and the JSPS KAKENHI Grants No. JP21H04880 and No. JP24K07485.}

\data{The data underlying the results of this paper are currently not publicly available but can be obtained from the authors upon reasonable request.}

\markboth{Appendix}{Appendix}

\appendix
\numberwithin{equation}{section}
\renewcommand{\thesection}{\Alph{section}}

\makeatletter
\renewcommand{\@seccntformat}[1]{Appendix \csname the#1\endcsname.\quad}
\makeatother

\section{Fidelity of quasi-ECS}
\label{sec:appendix_a}

In this Appendix we derive the analytical expression for the photon-subtracted TMSVS after propagation through a waveguide trimer and obtain an explicit analytical form for the fidelity in the lossless case.

The propagation of the quantized electromagnetic field in the coupled-waveguide system is governed by the Heisenberg equations of motion
\begin{align}
i\frac{\mathrm{d}}{\mathrm{d}z}
\begin{pmatrix}
\hat{a}^\dagger(z) \\
\hat{b}^\dagger(z) \\
\hat{c}^\dagger(z)
\end{pmatrix}
=
-
\begin{pmatrix}
\beta & \kappa & 0 \\
\kappa & \beta & \kappa \\
0 & \kappa & \beta
\end{pmatrix}
\begin{pmatrix}
\hat{a}^\dagger(z) \\
\hat{b}^\dagger(z) \\
\hat{c}^\dagger(z)
\end{pmatrix},
\label{eq:coupled_modes}
\end{align}
where $\kappa$ denotes the nearest-neighbour coupling coefficient, $\beta$ is the propagation constant of each of the waveguides, and $z$ is the longitudinal propagation coordinate. The operators $\hat{a}^\dagger$, $\hat{b}^\dagger$, and $\hat{c}^\dagger$ correspond to bosonic creation operators associated with waveguides $a$, $b$, and $c$, respectively.
The formal solution of Eq.~\eqref{eq:coupled_modes} yields the input--output relation
$\ket{\Psi(z)} = U(z)\ket{\Psi(0)}$,
with $\ket{\Psi(z)} = \big( \hat{a}^\dagger(z), \hat{b}^\dagger(z), \hat{c}^\dagger(z) \big)^{\mathrm{T}}$. The resulting evolution operator takes the form
\begin{align}
U(z)=
\begin{pmatrix}
\frac{1}{2}\!\left(1+\cos\Theta\right) &
-\frac{i}{\sqrt{2}}\sin\Theta &
\frac{1}{2}\!\left(\cos\Theta-1\right) \\
-\frac{i}{\sqrt{2}}\sin\Theta &
\cos\Theta &
-\frac{i}{\sqrt{2}}\sin\Theta \\
\frac{1}{2}\!\left(\cos\Theta-1\right) &
-\frac{i}{\sqrt{2}}\sin\Theta &
\frac{1}{2}\!\left(1+\cos\Theta\right)
\end{pmatrix},
\label{eq:evolution_operator}
\end{align}
where $\Theta = \sqrt{2}\,\kappa z$. In what follows, we set $\kappa=1$ without loss of generality.

The initial state consists of a TMSVS injected into modes $a$ and $c$, while mode $b$ is prepared in the vacuum state:
\begin{align}
\begin{split}
\ket{\psi_{\rm in}}
&=
(1-|r|^{2})^{1/2}
\sum_{l=0}^{\infty}
r^{l}
\ket{l}_{a}\ket{0}_{b}\ket{l}_{c}
\\
&=
(1-|r|^{2})^{1/2}
\sum_{l=0}^{\infty}
r^{l}
\frac{\big(\hat{a}^{\dagger}(0)\big)^{l}}{\sqrt{l!}}
\frac{\big(\hat{c}^{\dagger}(0)\big)^{l}}{\sqrt{l!}}
\ket{0}_{a}\ket{0}_{b}\ket{0}_{c}.
\end{split}
\end{align}
Under Heisenberg evolution, the creation operators transform as
\begin{align}
\hat{a}^{\dagger}(z) &=
U_{11}\hat{a}^{\dagger}(0)
+ U_{12}\hat{b}^{\dagger}(0)
+ U_{13}\hat{c}^{\dagger}(0),\\
\hat{c}^{\dagger}(z) &=
U_{31}\hat{a}^{\dagger}(0)
+ U_{32}\hat{b}^{\dagger}(0)
+ U_{33}\hat{c}^{\dagger}(0),
\end{align}
where $U_{ij}(z)$ are the matrix elements of the unitary evolution matrix.
The output state reads
\begin{align}\label{eq:Appendix_output}
\ket{\psi_{\rm out}}
=
\sqrt{1-|r|^{2}}
\sum_{l=0}^{\infty}
\frac{r^{l}}{l!}
\left(\hat{a}^{\dagger}(z)\right)^{l}
\left(\hat{c}^{\dagger}(z)\right)^{l}
\ket{0}_{a}\ket{0}_{b}\ket{0}_{c}.
\end{align}
Detection of $N$ photons in the central waveguide corresponds to the projector
\begin{align}
\hat{P}_{b}=\ket{N}_{b}\bra{N}_{b}.
\end{align}
The photon-subtracted state is obtained by expanding Eq.~\eqref{eq:Appendix_output} using the multinomial theorem and projecting onto the $N$-photon Fock subspace of mode $b$
\begin{align}\label{eq:photon_subtracted_appendix}
\begin{split}
\widetilde{\rho}_{\rm sub}
&=
\mathrm{Tr}_{b}
\left[
\hat{P}_{b}
\ket{\psi_{\rm out}}\bra{\psi_{\rm out}}
\right]
\\ & =
\sum_{v_{1},v_{3}=0}^{\infty}
\sum_{v_{1}',v_{3}'=0}^{\infty}
\sum_{l,l'=0}^{\infty}
\sum_{p_{1}, p_{2}, p_{3}=0}^{\infty}
\sum_{p'_{1}, p'_{2}, p'_{3}=0}^{\infty}
\mathcal{A}^{(N)}(z,r)
\ket{v_{1}}_{a}\ket{v_{3}}_{c}
\bra{v_{1}'}_{a}\bra{v_{3}'}_{c}.
\end{split}
\end{align}
where
\begin{align}\label{eq:A}
\begin{split}
\mathcal{A}^{(N)}(z,r)
&=
(1-|r|^{2})
\frac{r^{l} r^{l'}}{p_{1}!p_{2}!p_{3}! \, p_{1}'!p_{2}'!p_{3}'!}
\frac{l! \, l'!}{(v_{1}-p_{1})!(N-p_{2})!(v_{3}-p_{3})!
(v_{1}'-p_{1}')!(N-p_{2}')!(v_{3}'-p_{3}')!}
\\
&\quad\times
U_{11}^{p_{1}+v_{3}-p_{3}}
(U_{11}^{*})^{p_{1}'+v_{3}'-p_{3}'}
|U_{12}|^{2N}
U_{13}^{p_{3}+v_{1}-p_{1}}
(U_{13}^{*})^{p_{3}'+v_{1}'-p_{1}'}
N!\sqrt{v_{1}!v_{3}!v_{1}'!v_{3}'!}
\\
&\quad\times
\delta_{v_{1}+N+v_{3},\,2l}
\delta_{v_{1}'+N+v_{3}',\,2l'}
\delta_{p_{1}+p_2+p_{3},\,l}
\delta_{p'_{1}+p'_2+p'_{3},\,l'}.
\end{split}
\end{align}
The indices \(l\) and \(l'\) denote the photon-number indices in the ket and bra components of the input TMSVS, respectively. The indices \(v_1, v_3\) and \(v_1', v_3'\) represent the photon numbers in modes \(a\) and \(c\) in the ket and bra components of the conditional output state, respectively. The indices \(p_1, p_2, p_3\) and \(p_1', p_2', p_3'\) arise from the multinomial expansions of the evolved creation operators in the ket and bra components, respectively.
The normalized conditional state is therefore
\begin{align}
\hat{\rho}_{\rm sub}
=
\frac{\widetilde{\rho}_{\rm sub}}{\mathcal{N}_{\rm sub}},
\end{align}
$\mathcal{N}_{\rm sub}=\mathrm{Tr}\left(\widetilde{\rho}_{\rm sub}\right)$.
We now evaluate the fidelity of the normalized photon-subtracted state with respect to the odd entangled coherent state
\begin{align}
\ket{\psi_{\rm ECS}^{-}}
=
\mathcal{N}_{\alpha}^{-}
\left(
\ket{\alpha}_{a}\ket{\alpha}_{c}
-
\ket{-\alpha}_{a}\ket{-\alpha}_{c}
\right),
\end{align}
with normalization
\begin{align}
\mathcal{N}_{\alpha}^{-}
=
\left[2\left(1- e^{-4|\alpha|^{2}}\right)\right]^{-1/2}.
\end{align}
Using the Fock-basis expansion of coherent states
\begin{align}
\ket{\alpha}
=
e^{-|\alpha|^{2}/2}
\sum_{n=0}^{\infty}
\frac{\alpha^{n}}{\sqrt{n!}}\ket{n},
\end{align}
the fidelity is defined as
\begin{align}
F(\mathrm{ECS})
=
\bra{\psi_{\rm ECS}^{-}}
\hat{\rho}_{\rm sub}
\ket{\psi_{\rm ECS}^{-}}.
\end{align}
Substituting the Fock-basis expansion of the ECS and the expression for the conditional state, the fidelity takes the explicit form
\begin{align}
F(\mathrm{ECS})
&=
(\mathcal{N}_{\alpha}^{-})^{2}\frac{e^{-2|\alpha|^{2}}}{\mathcal{N}_{\rm sub}}
\sum_{v_{1},v_{3}=0}^{\infty}
\sum_{v_{1}',v_{3}'=0}^{\infty}
\frac{
\left[\alpha^{v_{1}+v_{3}}
- (-\alpha)^{v_{1}+v_{3}}\right]
\left[\alpha^{*(v_{1}'+v_{3}')}
- (-\alpha^{*})^{(v_{1}'+v_{3}')}\right]
}{
\sqrt{v_{1}!v_{3}!v_{1}'!v_{3}'!}
}
\mathcal{A}^{(N)}(z,r),
\end{align}
where $\mathcal{A}^{(N)}(z,r)$ is defined in Eq.~\eqref{eq:A}.

\section{Quasi-ECS under channel loss}
\label{sec:appendix_b}
In this Appendix we derive the analytical expression of the quasi-ECS after propagation through a lossy channel. The loss is modeled using a virtual beam splitter (BS) approach.
For an input Fock state $\ket{n}_1\ket{m}_2$, the BS transformation reads
\begin{align}\label{eq:beam_splitting_transformation_modes12}
\begin{split}
\ket{n}_1 \ket{m}_2 & \rightarrow
\frac{1}{\sqrt{n!\,m!}}
\sum_{k=0}^{n}
\sum_{l=0}^{m}
\binom{n}{k}\binom{m}{l}
(\sqrt{t})^{\,n+l-k}
(i\sqrt{1-t})^{\,m+k-l}
\sqrt{(m+n-k-l)!(k+l)!}\\
&\quad\times
\ket{m+n-k-l}_1
\ket{k+l}_2 ,
\end{split}
\end{align}
where $t$ is the BS transmission coefficient.
For the case $\ket{n}_1\ket{0}_2$, the transformation simplifies to
\begin{align}
\ket{n}_1\ket{0}_2
\rightarrow
\sum_{k=0}^{n}
\sqrt{\binom{n}{k}}\,
t^{(n-k)/2}
(i\sqrt{1-t})^{k}
\ket{n-k}_1
\ket{k}_2 .
\end{align}
The lossy state is then obtained by tracing out mode $2$:
\begin{align}
\hat{\rho}^{(\mathrm{lossy})}_1
=
\mathrm{Tr}_2\Big[
\hat{U}_{\mathrm{BS}}
(\hat{\rho}_1 \otimes \ket{0}_2\bra{0}_2)
\hat{U}_{\mathrm{BS}}^{\dagger}
\Big].
\end{align}
Accordingly, to model the effect of channel loss, we consider each mode interacting with a vacuum ancilla via a virtual BS of transmission coefficient $\eta$. The resulting transformation of the Fock-basis density matrix is then given by
\begin{align}\label{eq:density matrix transformation}
\ket{n}_{1}\!\bra{n'}_{1}
\rightarrow
\sum_{k=0}^{\min(n,n')}
\sqrt{\binom{n}{k}\binom{n'}{k}}
\,\eta^{(n+n'-2k)/2}
(1-\eta)^{k}
\ket{n-k}_{1}\!\bra{n'-k}_{1}.
\end{align}
where the environmental mode has been traced out. As a result, the system evolves into a mixed state.
Using Eq.~\eqref{eq:density matrix transformation}, the analytical expression of the quasi-ECS in Eq.~\eqref{eq:photon_subtracted_appendix} after channel loss becomes
\begin{align}
\widetilde{\rho}_{\rm lossy}
&=\frac{1}{\mathcal{N}_{\rm sub}}
\sum_{v_{1},v_{3}}
\sum_{v_{1}',v_{3}'}
\mathcal{A}^{(N)}(z,r)
\sum_{k_a=0}^{\min(v_1,v_1')}
\sum_{k_c=0}^{\min(v_3,v_3')}
\nonumber
\sqrt{
\binom{v_1}{k_a}
\binom{v_1'}{k_a}
\binom{v_3}{k_c}
\binom{v_3'}{k_c}
}
\eta^{(v_1+v_1'-2k_a)/2}
(1-\eta)^{k_a} \nonumber\\
&\quad\times
\eta^{(v_3+v_3'-2k_c)/2}
(1-\eta)^{k_c}
\ket{v_1-k_a}_a\bra{v_1'-k_a}_a
\otimes
\ket{v_3-k_c}_c\bra{v_3'-k_c}_c ,
\label{eq:rho_lossy_final}
\end{align}
where $\mathcal{A}^{(N)}(z,r)$ is defined in Eq.~\eqref{eq:A}. Taking care of the normalization factor
\begin{align}
    \hat{\rho}_{\rm lossy}=\frac{\widetilde{\rho}_{\rm lossy}}{\mathcal{N}_{\rm lossy}}, 
\end{align}  
where $\mathcal{N}_{\rm lossy}=Tr(\widetilde{\rho}_{\rm lossy})$. Because the photon-subtracted input is normalized and the pure-loss channel is trace preserving, $\mathcal{N}_{\rm lossy}=1$ analytically. The factor is retained here to make the normalization convention explicit.
%

\section{Purified Quasi-ECS}
\label{sec:appendix_c}
We now derive the analytical expression for the quasi-ECS after the purification operation introduced in Sec.~\ref{sec:loss_purification}. The two modes of the lossy state \(\hat{\rho}_{\rm lossy}\), denoted by \(a\) and \(c\), are each mixed with an ancillary single photon. The ancillary input modes are labeled \(a'_{\rm in}\) and \(c'_{\rm in}\), while the corresponding output modes are denoted by \(a'_{\rm out}\) and \(c'_{\rm out}\). The two directional couplers have the same transmission coefficient \(T\), and the total unitary is
\begin{align}
\hat{U}_{\rm tot}
=
\hat{U}_{a,a'}(T)
\otimes
\hat{U}_{c,c'}(T).
\label{eq:purification_total_unitary}
\end{align}
The purification event is heralded by detecting one photon in each ancillary output mode. Using the beam-splitter transformation in Eq.~\eqref{eq:beam_splitting_transformation_modes12}, we obtain
\begin{align}
{}_{a'_{\rm out}}\langle1|
\hat{U}_{a,a'}(T)
\ket{n}_{a}\ket{1}_{a'_{\rm in}}
=
\chi_n(T)\ket{n}_{a},
\label{eq:single_mode_catalysis}
\end{align}
where
\begin{align}
\chi_n(T)
=
T^{(n+1)/2}
-
n(1-T)T^{(n-1)/2}.
\label{eq:catalysis_coefficient}
\end{align}
The same relation applies to mode \(c\), with \(a\), \(a'_{\rm in}\), and \(a'_{\rm out}\) replaced by \(c\), \(c'_{\rm in}\), and \(c'_{\rm out}\), respectively.
Applying the two local catalysis operations to Eq.~\eqref{eq:rho_lossy_final}, the heralded purified state defined in Eq.~\eqref{eq:purified_state} is
\begin{align}
\widetilde{\rho}_{\rm purified}
&=\frac{1}{\mathcal{N}_{\rm sub}\mathcal{N}_{\rm lossy}}
\sum_{v_{1},v_{3}}
\sum_{v_{1}',v_{3}'}
\mathcal{A}^{(N)}(z,r)
\sum_{k_a=0}^{\min(v_1,v_1')}
\sum_{k_c=0}^{\min(v_3,v_3')}
\sqrt{
\binom{v_1}{k_a}
\binom{v_1'}{k_a}
\binom{v_3}{k_c}
\binom{v_3'}{k_c}
}
\nonumber\\
&\quad\times
\eta^{(v_1+v_1'-2k_a)/2}
(1-\eta)^{k_a}
\eta^{(v_3+v_3'-2k_c)/2}
(1-\eta)^{k_c}
\chi_{v_1-k_a}(T)
\chi_{v_1'-k_a}^{*}(T)
\chi_{v_3-k_c}(T)
\chi_{v_3'-k_c}^{*}(T)
\nonumber\\
&\quad\times
\ket{v_1-k_a}_{a}\bra{v_1'-k_a}_{a}
\otimes
\ket{v_3-k_c}_{c}\bra{v_3'-k_c}_{c},
\label{eq:rho_purified_final}
\end{align}
where \(\mathcal{A}^{(N)}(z,r)\) is defined in Eq.~\eqref{eq:A}, and \(\eta\) is the transmission coefficient of each lossy channel. The normalized purified state is given by
\begin{align}
    \hat{\rho}_{\rm purified}=\frac{\widetilde{\rho}_{\rm purified}}{\mathcal{P}_{\rm purified}}, 
\end{align}  
where $\mathcal{P}_{\rm purified}=Tr(\widetilde{\rho}_{\rm purified})$ denotes the success probability of the purification stage.

\section{Conditional Teleportation Output}
\label{sec:appendix_d}
We now derive the conditional output state of the teleportation protocol. Bob holds mode \(a\), while Alice holds mode \(c\). The unknown input state \(\ket{\psi_0}\) is prepared in mode \(d\),
\begin{align}
\ket{\psi_0}_{d}
=
\sum_{q=0}^{\infty}
\psi_q\ket{q}_{d},
\qquad
\hat{\rho}_0
=
\ket{\psi_0}_{d}\bra{\psi_0}_{d}.
\label{eq:teleportation_input_state}
\end{align}
Alice mixes modes \(c\) and \(d\) on a balanced beam splitter,
\begin{align}
\hat U_{c,d}
\equiv
\hat U_{c,d}(T_{\rm tel}),
\qquad
T_{\rm tel}=\frac{1}{2}.
\label{eq:teleportation_unitary}
\end{align}
The beam-splitter output modes are denoted again by \(c\) and \(d\). A measurement outcome \((x,y)\) corresponds to the detection of \(x\) photons in mode \(c\) and \(y\) photons in mode \(d\).
The conditional state in Bob's mode is
\begin{align}
\widetilde{\rho}^{(x,y)}_{\rm purified}
={}&
{}_{c}\langle x|
{}_{d}\langle y|
\hat U_{c,d}
\left[
\hat{\rho}_{\rm purified}
\otimes
\hat{\rho}_0
\right]
\hat U_{c,d}^{\dagger}
\ket{x}_{c}
\ket{y}_{d}.
\label{eq:conditional_teleportation_output}
\end{align}
Using Eq.~\eqref{eq:conditional_teleportation_output} together with Eq.~\eqref{eq:rho_purified_final}, this state is explicitly given by
\begin{align}
\widetilde{\rho}^{(x,y)}_{\rm purified}
&=\frac{1}{\mathcal{N}_{\rm sub}\mathcal{N}_{\rm lossy}\mathcal{P}_{\rm purified}}
\sum_{v_{1},v_{3}}
\sum_{v_{1}',v_{3}'}
\mathcal{A}^{(N)}(z,r)
\sum_{k_a=0}^{\min(v_1,v_1')}
\sum_{k_c=0}^{\min(v_3,v_3')}
\sqrt{
\binom{v_1}{k_a}
\binom{v_1'}{k_a}
\binom{v_3}{k_c}
\binom{v_3'}{k_c}
}
\nonumber\\
&\quad\times
\eta^{(v_1+v_1'-2k_a)/2}
(1-\eta)^{k_a}
\eta^{(v_3+v_3'-2k_c)/2}
(1-\eta)^{k_c}
\chi_{v_1-k_a}(T)
\chi_{v_1'-k_a}^{*}(T)
\chi_{v_3-k_c}(T)
\nonumber\\
&\quad\times
\chi_{v_3'-k_c}^{*}(T)
\sum_{q,q'=0}^{\infty}
\psi_q\psi_{q'}^{*}
\mathcal{B}_{v_3-k_c,q}^{(x,y)}
\mathcal{B}_{v_3'-k_c,q'}^{*(x,y)}
\ket{v_1-k_a}_{a}
\bra{v_1'-k_a}_{a},
\label{eq:conditional_teleportation_output_explicit}
\end{align}
where
\begin{align}
\mathcal{B}_{m,q}^{(x,y)}
={}&
{}_{c}\langle x|
{}_{d}\langle y|
\hat U_{c,d}
\ket{m}_{c}\ket{q}_{d}
\nonumber\\
={}&
\frac{\sqrt{x!y!}}{\sqrt{m!q!}}
\sum_{j=0}^{m}
\sum_{l=0}^{q}
\binom{m}{j}
\binom{q}{l}
\left(\frac{1}{\sqrt{2}}\right)^{m+l-j}
\left(\frac{i}{\sqrt{2}}\right)^{q+j-l}
\delta_{x,m+q-j-l}
\delta_{y,j+l}.
\label{eq:teleportation_bs_amplitude}
\end{align}
The operators in Eqs.~\eqref{eq:conditional_teleportation_output} and \eqref{eq:conditional_teleportation_output_explicit} are unnormalized after Alice's photon-counting measurement. For logical input $\ket{\psi_k}$ and accepted outcome $m$, the corrected output, its heralding probability and its normalized form are
\begin{align}\label{eq:normalized_corrected_output}
\widetilde\rho^{\rm corr}_{k,m} &= \hat C_m\widetilde\rho^{(k,m)}\hat C_m^\dagger,\nonumber\\
p_{k,m} &= \mathrm{Tr}(\widetilde\rho^{\rm corr}_{k,m}),\qquad
\hat\rho^{\rm corr}_{k,m} =\frac{\widetilde\rho^{\rm corr}_{k,m}}{p_{k,m}}.
\end{align}
The corresponding unpurified expressions follow from the same measurement projection applied to $\hat\rho_{\rm lossy}$, without the catalysis factors $\chi_n(T)$.

\bibliographystyle{iopart-num}
\bibliography{sample}

\end{document}